\documentclass[iop,apj,revtex4]{emulateapj}
\usepackage{apjfonts}
\usepackage{amsmath}
\usepackage{natbib}
\usepackage{graphicx}
\usepackage{hyperref}
\newcommand{\Hb}{H{$\beta$}}
\newcommand{\Ha}{H{$\alpha$}}
\newcommand{\msun}{\ensuremath{{\rm M}_{\odot}}}
\newcommand{\kms}{km\,s\ensuremath{^{-1}}} 
\newcommand{\mbh}{\ensuremath{M_\mathrm{BH}}}
\newcommand{\msigma}{\ensuremath{\mbh-\sigma_*}}
\newcommand{\sigmastar}{\ensuremath{\sigma_*}}
\newcommand{\sigmahb}{\ensuremath{\sigma_{\textrm{H}\beta}}}

\newcommand{\oiii}{[\ion{O}{3}]}
\newcommand{\oii}{[\ion{O}{2}]}
\newcommand{\ergs}{erg s\ensuremath{^{-1}}} 

\shortauthors{Oh et al.}

\begin{document}
\title{THE EXTENDED NARROW-LINE REGION OF TWO TYPE-I QUASI-STELLAR OBJECTS}

\author{Semyeong Oh$^{1}$},
\author{Jong-Hak Woo$^{1,7}$},
\author{Vardha N. Bennert$^{2}$},
\author{Bruno Jungwiert$^{3}$},
\author{Martin Haas$^{4}$},
\author{Christian Leipski$^{5}$},
\author{Marcus Albrecht$^{6}$}

\affil{$^{1}$Astronomy Program, Department of Physics and Astronomy, Seoul National University, Seoul, 151-742, Korea; smoh@astro.snu.ac.kr, woo@astro.snu.ac.kr}
\affil{$^{2}$Physics Department, California Polytechnic State University San Luis Obispo, CA 93407, USA; email: vbennert@calpoly.edu}
\affil{$^{3}$Astronomical Institute, Academy of Sciences of the Czech Republic,
Bocni II 1401/1a, CZ-141 31 Prague, Czech Republic; email: bruno@ig.cas.cz}
\affil{$^{4}$Astronomisches Institut, Ruhr-Universit{\"a}t Bochum,
Universit{\"a}tsstra{\ss}e 150, D-44801 Bochum, Germany; email: haas@astro.rub.de}
\affil{$^{5}$Max-Planck Institut f{\"u}r Astronomie,
K{\"o}nigstuhl 17, D-69117 Heidelberg, Germany; email: leipski@mpia-hd.mpg.de}
\affil{$^{6}$Argelander-Institut f\"ur Astronomie, Auf dem H\"ugel 71, 53121 Bonn, Germany; malbrecht@astro.uni-bonn.de}
\altaffiltext{7}{Author to whom correspondence should be addressed.}

\begin{abstract}

We investigate the narrow-line region (NLR) of two radio-quiet QSOs, 
PG1012+008 and PG1307+085, using high signal-to-noise spatially resolved long-slit spectra 
obtained with FORS1 at the Very Large Telescope. Although the emission is dominated 
by the point-spread function of the nuclear source,
we are able to detect extended NLR emission out to several kpc scales in both QSOs
by subtracting the scaled central spectrum from outer spectra.
In contrast to the nuclear spectrum, which shows a prominent
blue wing and a broad line profile of the \oiii\ line,
the extended emission reveals no clear signs of large scale outflows.
Exploiting the wide wavelength range, we determine the radial change of
the gas properties in the NLR, i.e., gas temperature, density, and ionization 
parameter, and compare them with those of Seyfert galaxies and type-II QSOs.
The QSOs have higher nuclear temperature and lower electron density than Seyferts, but 
show no significant difference compared to type-II QSOs,
while the ionization parameter decreases with radial distance, similar to
the case of Seyfert galaxies, 
For PG1012+008, we determine the stellar velocity dispersion of the host galaxy.
Combined with the black hole mass, we find that the luminous radio-quiet QSO
follows the local \msigma{} relation of active galactic nuclei.

\end{abstract}

\section{INTRODUCTION}
Since the discovery of correlations between the mass of black holes (BHs)
and the properties of their host-galaxy bulges 
\citep[][]{magorrian1998,ferrarese2000,gebhardt2000},
the interplay between active galactic nuclei (AGNs) ---
thought to represent a stage of galaxy evolution
in which BH is actively growing through accretion ---
and their host galaxies has received much attention.
In particular, AGN feedback seems to provide a promising way
to drive these relations by quenching both star-formation 
and accretion onto the BH, thus self-regulating BH growth
\citep[e.g.,][]{di-matteo2005,hopkins2006,sironi2010}.

Recently, there has been growing observational support for such a scenario
for various AGNs, e.g., radio-loud quasars \citep[e.g.,][]{nesvadba2008,fu2009},
broad-absorption line quasars \citep[e.g.,][]{crenshaw2003,moe2009},
local ultra-luminous infrared galaxies
\citep[e.g.,][]{fischer2010a,feruglio2010,sturm2011,rupke2011},
and high redshift AGNs \citep[e.g.,][]{tremonti2007,alexander2010,hainline2011}.
These observations are, however, often restricted to a handful of sources.
Moreover, it is difficult to determine the kinetic energy involved
and to prove that the observed outflows are indeed galaxy-scale 
radiatively-driven AGN outflows
(compared to outflows driven by e.g., star formation or jets).

One approach in the quest for observational signatures of AGN feedback is to
focus on the emission-line regions in the vicinity of the BH:
the so-called broad-line region (BLR) and narrow-line region (NLR),
which are photoionized by the central engine.
While BLRs are confined to sub-pc scales around an accretion disk,
producing kinematically broadened emission lines with
typical velocities of $10^{3}-10^{4}$~\kms{},
NLRs, characterized by narrow-lines with typical velocities of $10^{2}-10^{3}$~\kms{},
can span over kpc scales, which are comparable to the size of the bulge or 
even the entire galaxy \citep{boroson1984,stockton1987}.
Thus, as a direct interface between the AGN and its host galaxy, the NLR 
can provide important clues on the impact of BHs on their host galaxies 
and vice versa.

The NLR of Seyfert galaxies has been studied extensively over the last
decades \citep[e.g.,][]{mulchaey1996, ho1997, schmitt2003, falcke1998,
bennert2002, stoklasova2009}.
While most earlier studies have focused on the extent and morphology of the NLR,
the interplay between radio emission and NLR, and implications for the unified model of AGNs 
\citep[see][for review]{antonucci1993},
the goal of recent studies is to search for imprints of AGN feedback, e.g.,
in the form of outflowing gas \citep{schlesinger2009, muller-sanchez2011}.

In particular, the forbidden \oiii{} $\lambda$5007 (hereafter \oiii{}) emission line
is well known to show blue wings, which are generally interpreted as a
sign of outflow \citep[see][and references therein]{crenshaw2010}.
To study the kinematics of the ionized gas, spectroscopy of the extended NLR is essential.
However, such a study is especially challenging for
the most promising candidates to exhibit AGN feedback,
high-luminosity QSOs, due to the presence of their bright nuclei.

Recent studies have thus focused on type-II (obscured) QSOs
that have been discovered in large numbers from the Sloan Digital Sky Survey
\citep[SDSS;][]{zakamska2003, reyes2008}.
For example, \citet{greene2011} studied 15 luminous type-II QSOs at low-redshift
($z < 0.5$) with spatially resolved 
spectroscopy. Comparing the extent of the continuum and \oiii{} emission,
they argue that the AGN is ionizing the interstellar medium in the entire 
host galaxy. The large velocity dispersion
of this ionized gas out to kpc scales suggests that the gas is
disturbed on galactic scale. However, the velocity dispersion is below 500~\kms{} for most
objects in their sample, and the overall escaping fraction is less than 25\% with
a median of 2\%, even in the most extreme cases.
Similarly, \citet{villar-martin2011} report blue asymmetry of the \oiii{} emission line
in 11 objects out of a sample of 13 type-II quasars at $0.3 < z < 0.5$.
They interpret the asymmetry as a signature of outflows 
within a few kpc from the central engine. 
Furthermore, they find an anticorrelation between the degree by which
the kinematics of the outflow is perturbed
and its contribution to the total \oiii{} flux,
suggesting that only a small fraction of the total mass of the ionized
gas is involved in the outflow.

With this line of evidence for type-II quasars,  we here study the NLR of 
type-I QSOs, PG1012+008 and PG1307+085 with spatially resolved spectroscopy.
Benefiting from the wide spectral range and excellent seeing condition,
we investigate the kinematics and physical conditions of the NLR as a function of radial distance.
Throughout the paper, we assume a Hubble constant of $H_0$ =
70~\kms{}\,Mpc$^{-1}$, $\Omega_{\Lambda}$ = 0.7, and $\Omega_\textrm{M}$ = 0.3.

\begin{deluxetable*}{ccccccccccc}
\centering
\tablewidth{0.9\textwidth}
\tablecolumns{11}
\tablecaption{Observation} 
\tablehead{
\colhead{Name}&\colhead{$z$}&\colhead{Exposure}&\colhead{P.A.}&\colhead{Seeing}&
\colhead{S/N}&\colhead{Scale}&\colhead{$D_L$}&\colhead{$L_{[O III]}$}&\colhead{$L_B$} &\colhead{$R_e$}\\
\colhead{}&\colhead{}&\colhead{(s)}&\colhead{(deg)}&\colhead{(\arcsec)}&
\colhead{(pixel$^{-1}$)}&\colhead{(kpc\,arcsec$^{-1}$)}&\colhead{(Mpc)}&
\colhead{($\times 10^{42}$ \ergs)}&
\colhead{($\times 10^{45}$ \ergs})&\colhead{(\arcsec)}\\
\colhead{(1)}&\colhead{(2)}&\colhead{(3)}&\colhead{(4)}&\colhead{(5)}&\colhead{(6)}&
\colhead{(7)}&\colhead{(8)}&\colhead{(9)}&\colhead{(10)}&\colhead{(11)}
}
\startdata 
PG1012+008 & 0.187 & 7400 &  142 & $<0.7$ & 472--26   & 3.1 & 909.4 & 4.94 & 1.13& 3.4\\
PG1307+085 & 0.155 & 6800 &  112 & $<0.6$ & 507--5    & 2.7 & 739.2 & 4.53 & 1.41& 1.3
\enddata
\tablecomments{
Column 1. object name.
Column 2. redshift determined from \oii{} $\lambda$3727.
Column 3. total exposure time.
Column 4. position angle of the slit.
Column 5. seeing condition.
Column 6. signal-to-noise ratio at 5100 \AA{} from the center to the outermost aperture.
Column 7. angular diameter distance.
Column 8. luminosity distance.
Column 9. \oiii{} luminosity measured from \citet{bennert2002}.
Column 10. $B$-band luminosity from \citet{elvis1994}.
Column 11. effective radius derived from two-dimensional de Vaucouleurs fits 
to the surface brightness profile given in \citet{bahcall1997}.
}
\label{table1}
\end{deluxetable*}

\begin{figure}
\plotone{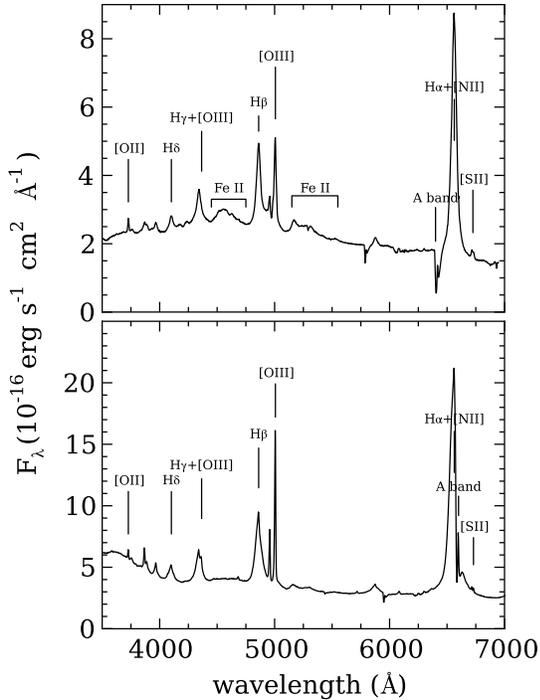}
\caption{Central de-redshifted spectra of PG1012+008 (top) and PG1307+085 (bottom).}
\label{f1}
\end{figure}

\section{OBSERVATIONS AND DATA REDUCTION}
\subsection{Sample Selection}

To study the NLR of luminous quasars, we started with the seven low-redshift
Palomar-Green (PG) quasars \citep{schmidt1983}, for which \oiii{} narrow-band imaging 
was obtained with the {\it Hubble Space Telescope} ({\it HST}) by \citet{bennert2002}.
Out of these seven objects, we selected those with the most extended NLR
that were visible from the Very Large Telescope (VLT), and at a redshift range
that allowed us to cover the broadest range of emission lines in optical spectra.
As a result, we selected two such objects, PG1012+008 ($z=0.187$) and PG1307+085 ($z=0.155$),
with an NLR size of 1\farcs59 ($\sim 5$~kpc) and 1\farcs39 ($\sim 3.7$~kpc),
respectively, based on the {\it HST} images \citep{bennert2002}.
The morphology of the \oiii{} emission is symmetric and compact in general \citep{leipski2006a}.
The surface brightness at the edge of the NLRs is $\sim 1.4 \times 10^{-15}$ \ergs{}cm$^{-2}$arcsec$^{-2}$ ($3-\sigma$ detection).

These two QSOs were also studied with the Wide-Field Camera on-board
{\it HST} \citep{bahcall1997}, and
with the Very-Large Array in the radio \citep{leipski2006a}.
Here, we briefly summarize the properties of the host galaxies and the radio emission of the 
sample based on these previous studies.
The optical image of PG1012+008 reveals a complex morphology:
It is undergoing a major merger with a companion galaxy at 
3\farcs3 ($\sim 10$~kpc) to the east, exhibiting clear tidal-tail structure.
Another smaller, compact galaxy at 6\farcs8 ($\sim 21$~kpc) north of the QSO
may also be interacting with the two galaxies.
In the case of PG1307+085, the host galaxy is an isolated small elliptical galaxy.

Both QSOs are radio-quiet with total fluxes less than 0.3 mJy at
8.4~GHz for PG1012+008 and 4.8~GHz for PG1307+085 \citep{leipski2006a}.
The isophotes of \oiii{} and the radio continuum show similar structure 
in both objects on a scale of $\sim 1$\arcsec{}, implying a radio-NLR interaction.  
Particularly for PG1012+008, a ``bending'' in the isophotes is apparent
in the \oiii{} and radio images in the SE direction,
although it is unclear what role is played by the merger.

The radial change of the \oiii{} line profile of both PG QSOs and its connection to the radio 
emission was investigated by \citet{leipski2006} based on the spectra obtained at
the New Technology Telescope at the European Southern Observatory (ESO).
Investigating the substructures of the \oiii{} line profile, they note 
that jet-gas interaction plays a role in the NLRs of radio-quiet quasars.
Although their method of analysis, i.e., spatially-resolved spectroscopy, is 
similar to ours,  
we present a more complete view on the NLRs using superior data with better 
spatial resolutions, wider wavelength ranges, and far higher signal-to-noise 
ratios (S/Ns), though we miss some kinematic subcomponents identified
due to relatively lower spectral resolution.

\subsection{Observations}

To investigate the properties of the NLR based on spatially resolved spectra,
we observed PG1012+008 and PG1307+085 with the visual and near-UV FOcal Reducer 
and low dispersion Spectrograph (FORS1) at the VLT
on 2005 April 1 and 3.
To achieve high spatial resolution, the observations were carried out
under good seeing conditions ($<0\farcs7$).
A slit width of 0\farcs7 was chosen to match the seeing. 
Based on the \oiii{} narrow-band images, we aligned the position angle of the slit
with the maximum radial extent of the \oiii{} emission. 
The spatial resolution is 0\farcs2\,pixel$^{-1}$.
We used grism 300V with an order sorting filter GG375, covering
a spectral range of 3300--7180~\AA{} with 2.6~\AA\,pixel$^{-1}$.
The instrumental broadening ($\sigma_{\textrm{inst}}$), 
determined from arc lamps and sky lines, is 216~\kms{} at 5000~\AA.
For flux calibrations, we observed a spectrophotometric standard star G60-54.
Table~\ref{table1} summarizes the observations.

\subsection{Data Reduction}

Standard data reduction, i.e., bias subtraction, flat-fielding, wavelength
calibration, flux calibration, and atmospheric absorption correction was carried out 
using a series of IRAF scripts developed for long-slit spectroscopy 
\citep[e.g.,][]{woo2005, woo2006}. 
Cosmic rays were removed using L.A.Cosmic \citep{van-dokkum2001}.
We performed wavelength calibration using arc lines and obtained      
a spectral resolution of 2.6~\AA\,pixel$^{-1}$ with 
standard deviation of $\sim0.25$~\AA{}. 

We extract seven spectra along the slit (spatial direction),
covering $\pm3\farcs6$.
Positive direction indicates SE and NW for PG1012+008 and PG1307+085, respectively.
The aperture size was increased from 0\farcs6 at the center
to 1\farcs8 at the outer radii by a step of 0\farcs4,
in order to boost the S/N.  
Figure~\ref{f1} presents the spectra extracted from the central aperture 
for both targets. 
Note that the region near \Ha{} suffers from the atmospheric absorption ($A$ band).
We attempted a correction using the standard star spectra.
However, we were unable to measure the [\ion{N}{2}] and 
narrow \Ha{} emission lines, due to the dominance of the broad
\Ha{} line.

\begin{figure}
\plotone{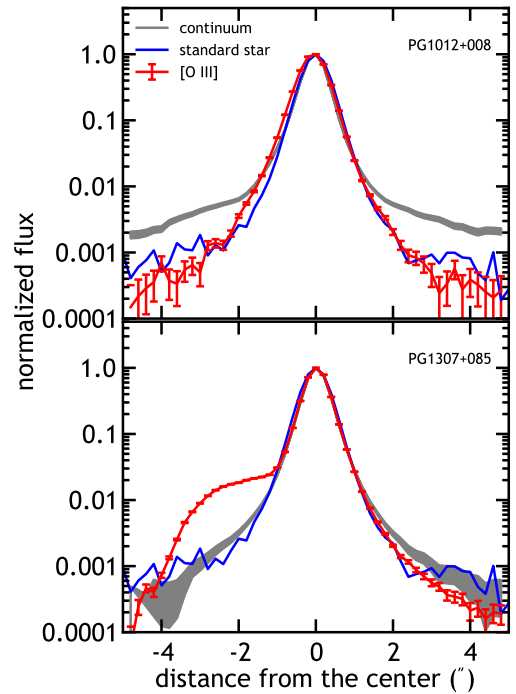}
\caption{Normalized spatial profile of continuum, \oiii{} emission line, and
the standard star, respectively for PG1012+008 (top) and PG1307+085 (bottom).
For the continuum,  the shaded region indicates 3$\sigma$ errors.}
\label{f2}
\end{figure}

\begin{figure*}
\plottwo{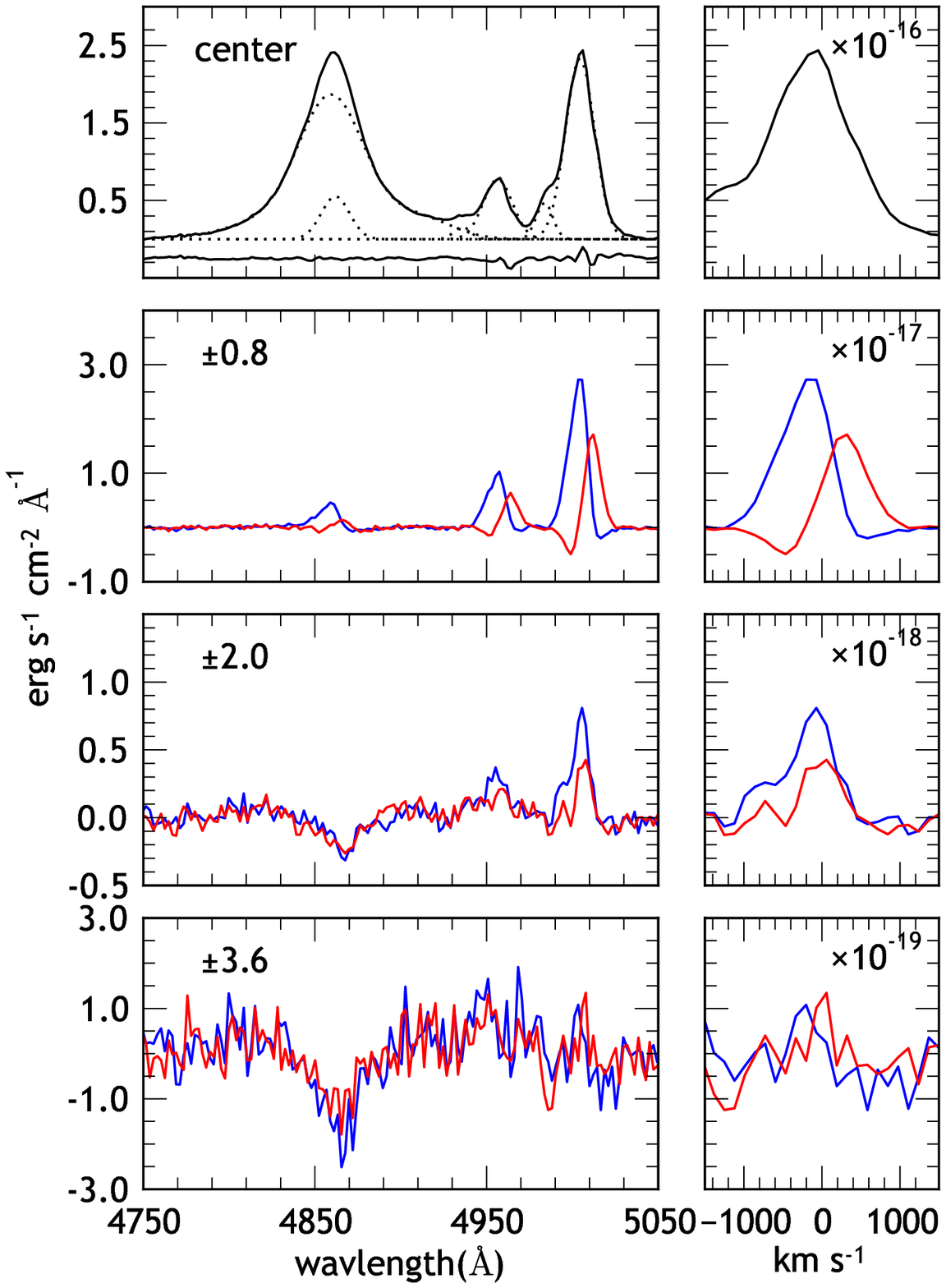}{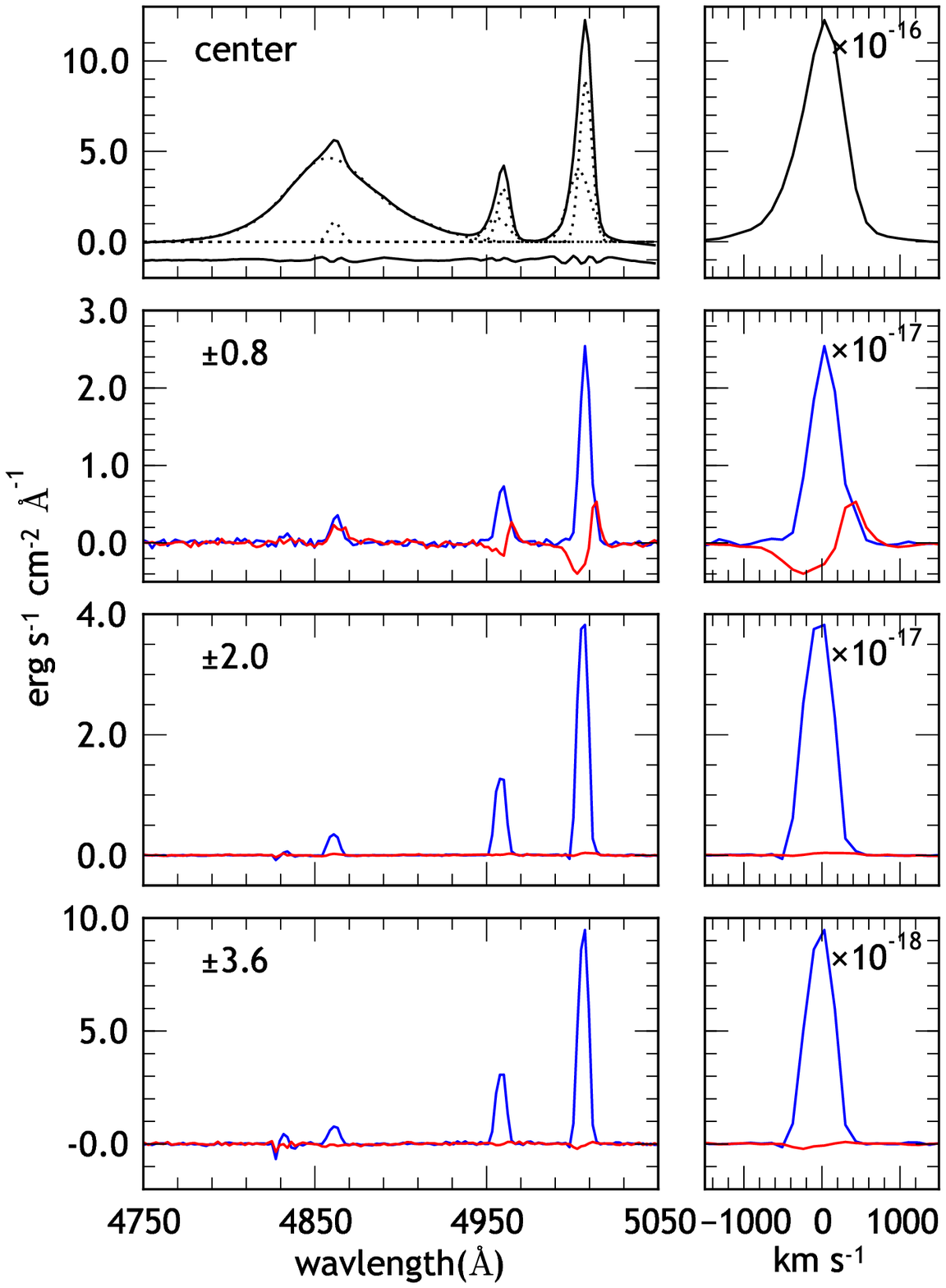}
\caption{Continuum-subtracted central spectrum 
and residuals after subtracting a scaled central spectrum
from the outer spectra, respectively, for PG1012+008 (left)
and PG1307+085 (right). Blue and red lines correspond to the minus and plus side
of the center along the slit, respectively, and the number on the upper left
corresponds to the distance from the center in units of arcseconds.
The best-fit models using several subcomponents are represented by dashed lines
(see the text for details), the residuals are presented with an arbitrary offset.
The \oiii{} emission line is enlarged with
the wavelength axis converted to velocity in the right panels.
The reference redshift is determined from \oii{} $\lambda3727$ line in
the central spectrum.
}
\label{f3}
\end{figure*}

\section{RESULTS}

\subsection{Extended Narrow-line Region}
\label{sec:vdisp}
We analyze spatially resolved spectra to investigate the spatial extension of the NLR.
First, to compare the relative strength of the emission 
between the central and extended regions, we determine the spatial profiles of 
\oiii{}, continuum, and standard star, respectively.
In the case of the continuum, which is dominated by the AGN and thus
expected to be spatially broadened by the seeing effect,
we collapse the spectra in the continuum wavelength range 5080--5120~\AA{} to determine
the spatial profile.
For \oiii{}, we sum the flux in a spectral range of 
$\sim 30$\AA{} centered on the peak of \oiii{},
and subtract the continuum level previously determined.
We also determine the point-spread function (PSF) based on the spectral 
image of the standard star.
Figure 2 presents the spatial profiles of both targets, which are normalized by 
their maximum flux value.
In the case of PG1012+008, the continuum spatial profile is consistent with the PSF
in the inner region at -1\arcsec{} $< r < 2$\arcsec{}, 
while at the outer region, the contribution of the host galaxy star light becomes
dominant as indicated by a broad extension in the spatial profile.
Although the spatial profile of \oiii{} is dominated by the PSF,
the left side of the profile is slightly extended over the PSF or the continuum 
spatial profile.
In the case of PG1307+085, we detect an asymmetric profile of \oiii{} with a distinct
bump at $-3$\arcsec, representing an extended NLR.
Thus, we conclude that the \oiii{} NLR in both objects is marginally detected.

To further test whether the \oiii{} emission in the off-center region is consistent with
being the PSF wings of the \oiii{} emission from the center,
we use the central spectrum of each object as a template
to subtract the seeing-scattered broad and narrow lines from the outer spectra.
For generating a template of the central spectrum, we fit the continuum and the broad component 
of \Hb{} (see the top panel of Figure 3), then combine the best fit models with the profiles of 
the narrow \Hb\ and \oiii{}. To account for the contribution from the host galaxy star light,
we also add a linear background level.

Figure~\ref{f3} presents the continuum-subtracted central spectrum (top panel) and 
residual spectra of outer region after subtracting a scaled central spectrum (middle and bottom
panels). 
The strong broad \Hb{} component observed at the outer spectra is well subtracted along with 
the contribution of the PSF wings of narrow emission lines from the center,
leaving a flat continuum and residual narrow emission lines.
These residual \oiii{} and \Hb{} lines represent the  extended NLR in both objects.
As expected from the investigation of spatial profiles (Figure~\ref{f2}),
both objects show very different spectra as the distance from
the center increases.
For PG1012+008, the extended \oiii{} emission is detected
out to $r \sim 2$\arcsec{} with similar flux on either side, while
PG1307+085 shows the extended emission only on one side.
The host galaxy starlight, e.g., \Hb\ absorption, starts to show at $r \gtrsim 2$\arcsec{}
for PG1012+008, while it is not visible in PG1307+085, presumably 
due to a higher contrast of the nuclear AGN luminosity 
to the host galaxy star light.
Thus, we detect the extended NLR emission out to $r > 2.0 \pm 0\farcs7$ ($6 \pm 2$~kpc), 
which is larger than the sizes measured by \citet{bennert2002} since the depth of our 
observation is roughly 20 times larger.

The central spectrum of both targets shows an asymmetric \oiii{} line profile
with prominent blue wings, which can be interpreted as outflows.
To properly measure the velocity of \oiii{}, we model the \Hb{}--\oiii{} region of 
the central spectrum as follows:
First, we use a power law + IZwI template from \citet{boroson1992}
to fit strong \ion{Fe}{2} multiplets in the central spectrum of PG1012+008.
For PG1307+085, the \ion{Fe}{2} emission is an order of magnitude weaker 
(compared to the continuum) and considered negligible.
Then, we model the \oiii{} line profile with two Gaussians along with broad and narrow \Hb{}.
We use Gauss-Hermite polynomials of order six for the broad component of \Hb{},
and the ratio of the two \oiii{} lines is fixed to the theoretical value (1/3).
The best-fit models of the continuum-subtracted central spectra
are shown in the top panels of Figure~\ref{f3}.

The ``blue'' component of \oiii{}, presumably representing the outflowing gas,
is blueshifted from the reference redshift by $1380$~\kms{} and
$173$~\kms{} for PG1012+008 and PG1307+085, respectively.
The full width at half maximum (FWHM) velocity of the total line profile
measured from the central spectra is $1050$~\kms{} for PG1012+008 and
$540$~\kms{} for PG1307+085, respectively.
In contrast, as shown in Figure 3, the extended \oiii{} emission is narrower 
than the \oiii{} in the central spectrum for both objects, and 
most of the line flux is coming from low velocity gas ($V < 500$~\kms).
This indicates that the outflowing component with $V > 500$~\kms{}
is unresolved at the spatial resolution of our current observation
and confined to within 0\arcsec7 scale ($\sim$2~kpc).
Thus, the emission-line profile of the extended gas indicates that 
the extended gas is kinematically quiescent in both QSOs.
We conclude that while we detect extended emission
out to $r \approx 2\arcsec{}$ (6~kpc),
there are no signs of a large-scale($\sim 10$~kpc) outflow.

We briefly compare our results with those by \citet{leipski2006}.
The red asymmetry towards the south of the nucleus in PG1012+008 detected by 
\citet{leipski2006} may correspond to the redshifted residual \oiii{} line
at 0\farcs8.
We miss some of the complex velocity structure seen by \citet{leipski2006}
due to lower spectral resolution.
As noted by \citet{leipski2006}, PG1307+085 shows a dramatic change in line profile
especially to the SE (``minus'' in this study) side.
The residual narrow line to the SE of PG1307+08 is unresolved at our spectral
resolution. \citet{leipski2006} report a velocity dispersion of $\sim 69$~\kms{}
for this component, which is consistent with our results.
While our results are in overall agreement with those presented in \citet{leipski2006}, 
given the differences in spectral resolution and S/N,
we find that the broad \oiii{} line profile of PG1307+085 towards the south
is consistent with being a PSF wing rather than true physical
extension of outflowing gas. 

The \oiii{}/\Hb{} line ratio is a diagnostic of the ionization
mechanism \citep{baldwin1981}.
In the case of PG1012+008, we cannot rule out that the ionization
of the extended gas is at least partially due to star formation.
For PG1307+085, however,
the high ($>5$) \oiii{}/\Hb{} ratio at $r \leq -0\farcs8$ from the center 
suggests that the QSO is the main source of ionization.

\begin{figure}
\centering
\plotone{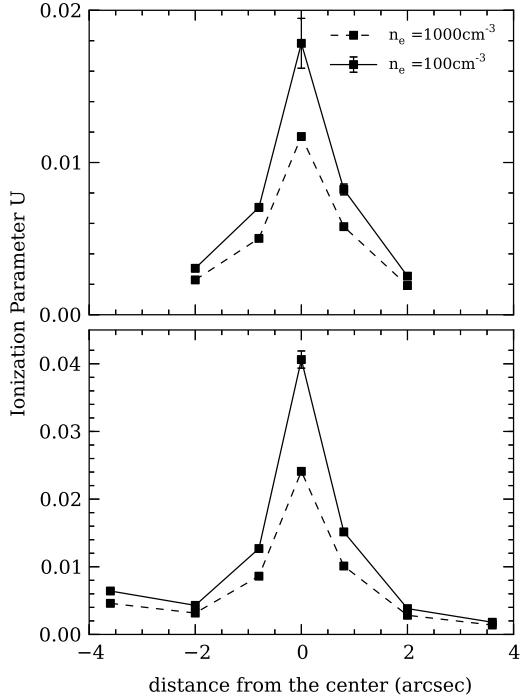}
\caption{Ionization parameter $U$ as a function of radial distance for PG1012+008 (top) and
PG1307+085 (bottom).}
\label{f6}
\end{figure}

\begin{deluxetable*}{ccccccc}
\centering
\tablecaption{Black Hole Masses}
\tablehead{
\colhead{Name} &
\colhead{$L_{\textrm{bol}}/L_{\textrm{Edd}}$\tablenotemark{a}} &
\colhead{$\log L_{5100}$} &
\colhead{$\log L_{5100}$\tablenotemark{b}} &
\colhead{$\sigma_{\textrm{\Hb{}}}$} &
\colhead{$\log{\mbh /\textrm{\msun{}}}$} &
\colhead{\sigmastar{}\tablenotemark{c}} \\
\colhead{} &
\colhead{} &
\colhead{(\ergs)} &
\colhead{(\ergs)} &
\colhead{(\kms{})} &
\colhead{} &
\colhead{(\kms)} } 
\startdata
PG1012+008 & 0.16 & 44.3 & 45.0 & 1729 & 8.42 & 196$\pm$30 \\
PG1307+085 & 0.14 & 44.2 & 45.0 & 1885 & 8.50 &
\enddata
\tablenotetext{a}{The bolometric correction factor for $L_{5100}$ is taken from \citet{netzer2009}.}
\tablenotetext{b}{From \citet{vestergaard2006}.}
\tablenotetext{c}{Measured at 2\arcsec{} from the center.}
\label{table2}
\end{deluxetable*}

\begin{figure}
\plotone{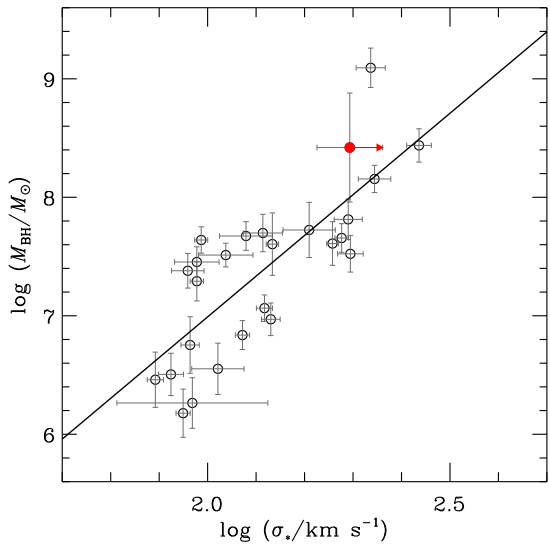}
\caption{PG1012+008 on the \msigma{} relation. Black data and fit are taken
from \citet{woo2010}. Our target is marked as a red circle with the arrow indicating that
the stellar velocity dispersion measured at $\pm2$\arcsec{} is a lower limit to the true
luminosity-weighted bulge velocity dispersion.}
\label{f7}
\end{figure}

\subsection{Physical Conditions}
In this section, we measure temperature, density, and ionization parameter of 
our targets to compare them with other AGNs.

\subsubsection{Temperature and Density}
We determined the temperature of the NLR gas using
the \oiii{} ($\lambda$4959 + $\lambda$5007)/$\lambda$4363 ratio 
as outlined in \citet{osterbrock1989}:

\begin{equation}
\textrm{\oiii{}} (\frac{\lambda4959 + \lambda5007}{\lambda4363}) 
\simeq \frac{7.73 \exp{(3.29\times10^4/T)}}{1+4.5\times10^{-4}(n_e/T^{1/2})},
\end{equation}
where, $n_e$ is electron density and $T$ is gas temperature.
With typical values of $n_e$ = $10^{3}-10^{4}$~cm$^{-3}$ and $T = 10^4$K,
the dependence on $n_e$ is negligible.

For PG1307+085, we were able to measure temperature and density using
\oiii{} and [\ion{S}{2}] line ratios from the center towards the south. 
The derived temperature is 25500 $\pm$ 1480K at the center,
and decreases with radial distance to 13,400 $\pm$ 11,700K at 3\farcs6 ($\sim$9.7~kpc).
Using the same methods, \citet{bennert2006Sy1,bennert2006Sy2} estimated the
average temperature of four Seyfert 1 and four Seyfert 2 galaxies in the nuclear region as 33,590 $\pm$ 7070K and 14,470 $\pm$ 440K, respectively, while  
\citet{greene2011} estimated a temperature range of $T = 11,000--23,000$ K for 15 type-II QSOs.
Note that the measurements of the Seyfert galaxies 
in \citet{bennert2006Sy1,bennert2006Sy2} included reddening correction
determined from the Balmer decrement, while \citet{greene2011} and this work did not
correct for reddening.
Note also that the physical scale in each measurements differs: for the Seyfert
galaxies it corresponds to a few pc in the center, 
while for the type-II QSOs in \citet{greene2011},
it corresponds to 3\arcsec{} fiber aperture of SDSS spectra,
covering $\sim 11$~kpc at the median redshift.
Although a direct comparison is not straightforward,
the type-I QSO PG1307+085 reaches lower
temperature at the center compared to Seyfert 1 galaxies
and similar to type-II QSOs.

The density was measured in a standard way from the [\ion{S}{2}] $\lambda$6716/6731 ratio using
IRAF {\it temden} task,\footnote{http://stsdas.stsci.edu/nebular/temden.html}
correcting for the effect of temperature using the derived values.
The derived density ranges between 584 and 169 cm$^{-3}$, decreasing with radial distance (0\arcsec--2\farcs0).
These values are similar to the mean density $n_e$ = 335 cm$^{-3}$ of type-II QSOs 
estimated by \citet{greene2011},
and slightly lower than the mean nuclear density of Seyfert galaxies: 
1070 $\pm$ 180 cm$^{-3}$ for Seyfert 1 galaxies \citep{bennert2006Sy1}
and 1100 $\pm$ 315 cm$^{-3}$ for Seyfert 2 galaxies \citep{bennert2006Sy2}.
However, one should note that the nuclear density varies in a wide range
($\sim$ 300--2500 cm$^{-3}$) among Seyfert 1 and 2 galaxies studied 
by \citet{bennert2006Sy1, bennert2006Sy2},
and it is not yet clear if the difference is significant.
The higher and denser NLR of type-I QSOs can be a natural result
of a dusty torus blocking the central region in type-II QSOs.

\subsubsection{Ionization Parameter}

The ionization parameter is defined by the ratio of ionizing density to
hydrogen density as $U = Q(H) / 4 \pi c r^2 n_e$, where $Q(H)$ is 
the number of the ionizing photons emitted by the AGN per second
and $n_e$ is electron density.
By measuring the emission-line flux ratio \oii{} $\lambda3727$/\oiii{} $\lambda5007$
the ionization parameter can be empirically derived \citep{penston1990}. 
We estimate the ionization parameter as a function of radial distance
for both targets assuming $n_e=100$ and $n_e=1000$, respectively,
based on the calculations provided by \citet[][; Figure~\ref{f6}]{komossa1997}. 

The ionization parameter peaks at the center, and decreases with radius 
for both of our targets.
Given that the density decreases radially as well,
the actual decrease of the ionization parameter is slower than shown in 
either one of the curves (solid versus dashed);
more accurately, it would start in the center on the dashed line and then move
toward values on the solid line with increasing distance from the center.

Note that we did not apply any reddening correction to the line fluxes.
If the radial decrease of the ionization parameter were entirely due to change 
in reddening, a simple calculation using the standard 
extinction curve shows that $\triangle{}E(B-V) \sim 2$ would be required 
between the center and the outermost apertures.
Considering that the average global reddening of Seyfert 1 galaxies is 
$E(B-V)\sim0.4$ \citep{bennert2006Sy1},
it is unlikely that the radial trend is purely caused by reddening,
although there might be a small contribution from reddening.

Qualitatively, the decrease of the ionization parameter with radius is consistent with
the radial trend found in Seyfert galaxies \citep{bennert2006Sy1,bennert2006Sy2}.
The average nuclear ionization parameter of the two targets
is $U_{n_e = 1000{\textrm cm}^{-3}} = 1.79\times10^{-2}$, which is approximately 
five times larger than Seyfert 1 and 2 galaxies \citep{bennert2006Sy1,bennert2006Sy2}.

\subsection{Black Hole Masses and Stellar-velocity Dispersion}
We estimate BH masses from the velocity and size of the BLR based on
virial assumptions \citep[e.g.,][]{woo2002, park2012}. 
In practice, we use the single-epoch virial mass
estimator using the AGN continuum luminosity at 5100 \AA\ ($L_{5100}$)
and the line dispersion of the broad \Hb\ line \sigmahb.
From the nuclear spectra extracted with an aperture of 2\arcsec{} radius,
including $>99\%$ of the light, we measure $L_{5100}$ and \sigmahb{}, 
after subtracting \ion{Fe}{2}  contamination.  
Compared to the values from \citet{vestergaard2006}, our $L_{5100}$ is 0.77 dex smaller
probably due to slit losses. Therefore, we use $L_{5100}$ from \citet{vestergaard2006}.

Assuming a virial factor of $\log{f} = 0.72$ \citep{woo2010}, we calculate \mbh{}.
Recently, \citet{park2012} reported the systematic difference of
the \Hb{} line profile between the single-epoch and the rms spectra.
Correcting for this effect using Equation (13) in \citet{park2012},
the mass estimates of \mbh{} are decreased by $\sim 0.18$~dex (Table~\ref{table2}).
The estimates are consistent with those given in \citet{vestergaard2006} within the
errors of single-epoch BH mass ($\sim 0.46$~dex).

For PG1012+008, we can also measure stellar-velocity dispersion since
we clearly detect the host galaxy (at 2\arcsec{} from the center).
Due to the combined effect of the contamination from the AGN continuum and
emission lines, and the presence of 4000~\AA{} break, we consider the measurement 
based on the Ca H+K absorption lines as uncertain. In the case of the $G$-band region, 
broad H$\gamma$ affects the fitting. 
Thus, stellar-velocity dispersion is determined
using the spectral range 5030--5300, including
several strong stellar lines such as the \ion{Mg}{1} b triplet (5172~\AA) and Fe (5270~\AA)
\citep[e.g.,][]{woo2006,woo2008,greene2006}.

Both measurements now enable us to place the PG1012+008 on the
\msigma{} relation in Figure~\ref{f7}, where the 24 reverberation-mapped AGNs and the best-fit 
\msigma{} relation from \citet{woo2010} are overplotted for comparison.
Note that the stellar velocity dispersion of PG1012+008 is measured at 2\arcsec{} from the center,
and can thus be considered a lower limit for the luminosity-weighted velocity dispersion
of the bulge since the velocity dispersion generally increases toward the center
\citep[e.g.,][]{kang2013}.
Even when taking into account an expected shift to the right,
the object falls onto the local \msigma{} relation (Figure~\ref{f7}), indicating that
at least this low-$z$ high-luminosity QSO is following the same \msigma{}
relation.

\section{DISCUSSION AND SUMMARY}

We investigated the kinematics and physical condition of the NLR in two type-I QSOs,
PG1012+008 and PG1307+085, using spatially resolved spectroscopy.
The emission from the NLRs of these two targets is dominated by a
central point source
with slight deviation from the expected PSFs. 
However, modeling the \Hb--\oiii{} region of the outer spectra
clearly reveals the spatially extended weak NLR out to $r \approx 2\arcsec{}$ (6~kpc)
from the center, spanning a larger extent than previously
reported from {\it HST} \oiii{} images by \citet{leipski2006} due to the increase in depth
of our observations.
The \oiii{} line of the nuclear spectrum of both targets shows
blue wings and broad widths with FWHM velocities exceeding $500$~\kms{}.
However, the same feature is not present in the outer spectra,
indicating that the outflow is confined within the PSF size ($\sim$2~kpc) and does not 
extend to galactic scales.

Compared to similar studies of type-II QSOs \citep{humphrey2010, greene2011},
the NLR of the two type-I QSOs studied here appears to be less extended.
Due to the small sample size, and the intrinsic difficulty
of inferring the true ``size'' of emission in long-slit observations,
we are unable to draw a conclusion on whether
type-I QSOs actually have less extended NLRs.
Recently, \citet{husemann2013} carried out an integral-field spectroscopy
of a large sample of nearby type-I QSOs revealing a typical size of 10~kpc,
similar to that of type-II QSOs studied by \citet{greene2011}.
They also caution that the difference between the size
determined by long-slit spectroscopy and the size based on a two-dimensional spectroscopy
is on average a factor of two, and can be as high as a factor of six.
Even so, the mean size of NLRs in type-I QSOs determined by \citet{husemann2013}
seems to be smaller than that of type-IIs (28~kpc) based on a comparable integral 
field spectroscopy by \citet{liu2013a}, although these type-II QSOs are 
at a slightly higher redshift.

The lack of galactic scale outflows in type-I QSOs has also been
pointed out by \citet{husemann2013}. Out of 31 type-I QSOs in their study,
only three show kpc-scale outflow, and the spectra of extended emission
generally show narrower \oiii{} line profile than the central QSO spectrum.
In the framework of the unified model of AGNs \citep{antonucci1993},
while the outflow in the projected plane may look less extended in type-Is than type-IIs,
the projected velocity is expected to be larger as we look directly
into the ionization cone. However, it is not yet clear whether type-Is have larger 
projected outflow velocities than type-IIs \citep[see][]{husemann2013, liu2013a}.  
At the same time, although the \oiii{} emission is expected to be more
elongated or biconical for type-II QSOs,
the morphology of \oiii{} emission seems to be more
elongated in type-I QSOs \citep{husemann2013}.
It is possible that in a merger-driven evolutionary scenario \citep[e.g.,][]{sanders1988},
AGN feedback in the ionized gas occurs mainly in obscured (type-II) phase.
The sample probed here may still not be quite luminous enough 
to drive galactic wide energetic outflows that expel gas from the host galaxy
and quench star formation
as those found at high-redshift \citep[e.g.,][]{cano-diaz2012}.
Solid evidence for large-scale outflows related to AGN feedback still remains elusive.

\acknowledgments
We thank the anonymous referee for constructive suggestions, which improved the manuscript.
This work has been supported by the Basic Science Research Program
through the National Research Foundation of Korea funded
by the Ministry of Education, Science and Technology (2012-006087). 
M.H. is supported by the Nordrhein-Westf\"alische
Akademie der Wissenschaften und der K\"unste, 
funded by the Federal Republic of Germany and the 
state Nordrhein-Westfalen.


\begin{thebibliography}{61}
\expandafter\ifx\csname natexlab\endcsname\relax\def\natexlab#1{#1}\fi

\bibitem[{{Alexander} {et~al.}(2010){Alexander}, {Swinbank}, {Smail},
  {McDermid}, \& {Nesvadba}}]{alexander2010}
{Alexander}, D.~M., {Swinbank}, A.~M., {Smail}, I., {McDermid}, R., \&
  {Nesvadba}, N.~P.~H. 2010, \mnras, 402, 2211

\bibitem[{{Antonucci}(1993)}]{antonucci1993}
{Antonucci}, R. 1993, \araa, 31, 473

\bibitem[{{Bahcall} {et~al.}(1997){Bahcall}, {Kirhakos}, {Saxe}, \&
  {Schneider}}]{bahcall1997}
{Bahcall}, J.~N., {Kirhakos}, S., {Saxe}, D.~H., \& {Schneider}, D.~P. 1997,
  \apj, 479, 642

\bibitem[{{Baldwin} {et~al.}(1981){Baldwin}, {Phillips}, \&
  {Terlevich}}]{baldwin1981}
{Baldwin}, J.~A., {Phillips}, M.~M., \& {Terlevich}, R. 1981, \pasp, 93, 5

\bibitem[{{Bennert} {et~al.}(2002){Bennert}, {Falcke}, {Schulz}, {Wilson}, \&
  {Wills}}]{bennert2002}
{Bennert}, N., {Falcke}, H., {Schulz}, H., {Wilson}, A.~S., \& {Wills}, B.~J.
  2002, \apjl, 574, L105

\bibitem[{{Bennert} {et~al.}(2006{\natexlab{a}}){Bennert}, {Jungwiert},
  {Komossa}, {Haas}, \& {Chini}}]{bennert2006Sy1}
{Bennert}, N., {Jungwiert}, B., {Komossa}, S., {Haas}, M., \& {Chini}, R.
  2006{\natexlab{a}}, \aap, 459, 55

\bibitem[{{Bennert} {et~al.}(2006{\natexlab{b}}){Bennert}, {Jungwiert},
  {Komossa}, {Haas}, \& {Chini}}]{bennert2006Sy2}
---. 2006{\natexlab{b}}, \aap, 456, 953

\bibitem[{{Boroson} \& {Green}(1992)}]{boroson1992}
{Boroson}, T.~A., \& {Green}, R.~F. 1992, \apjs, 80, 109

\bibitem[{{Boroson} \& {Oke}(1984)}]{boroson1984}
{Boroson}, T.~A., \& {Oke}, J.~B. 1984, \apj, 281, 535

\bibitem[{{Cano-D{\'{\i}}az} {et~al.}(2012){Cano-D{\'{\i}}az}, {Maiolino},
  {Marconi}, {Netzer}, {Shemmer}, \& {Cresci}}]{cano-diaz2012}
{Cano-D{\'{\i}}az}, M., {Maiolino}, R., {Marconi}, A., {et~al.} 2012, \aap,
  537, L8

\bibitem[{{Crenshaw} {et~al.}(2003){Crenshaw}, {Kraemer}, \&
  {George}}]{crenshaw2003}
{Crenshaw}, D.~M., {Kraemer}, S.~B., \& {George}, I.~M. 2003, \araa, 41, 117

\bibitem[{{Crenshaw} {et~al.}(2010){Crenshaw}, {Schmitt}, {Kraemer},
  {Mushotzky}, \& {Dunn}}]{crenshaw2010}
{Crenshaw}, D.~M., {Schmitt}, H.~R., {Kraemer}, S.~B., {Mushotzky}, R.~F., \&
  {Dunn}, J.~P. 2010, \apj, 708, 419

\bibitem[{{Di Matteo} {et~al.}(2005){Di Matteo}, {Springel}, \&
  {Hernquist}}]{di-matteo2005}
{Di Matteo}, T., {Springel}, V., \& {Hernquist}, L. 2005, \nat, 433, 604

\bibitem[{{Elvis} {et~al.}(1994){Elvis}, {Wilkes}, {McDowell}, {Green},
  {Bechtold}, {Willner}, {Oey}, {Polomski}, \& {Cutri}}]{elvis1994}
{Elvis}, M., {Wilkes}, B.~J., {McDowell}, J.~C., {et~al.} 1994, \apjs, 95, 1

\bibitem[{{Falcke} {et~al.}(1998){Falcke}, {Wilson}, \& {Simpson}}]{falcke1998}
{Falcke}, H., {Wilson}, A.~S., \& {Simpson}, C. 1998, \apj, 502, 199

\bibitem[{{Ferrarese} \& {Merritt}(2000)}]{ferrarese2000}
{Ferrarese}, L., \& {Merritt}, D. 2000, \apjl, 539, L9

\bibitem[{{Feruglio} {et~al.}(2010){Feruglio}, {Maiolino}, {Piconcelli},
  {Menci}, {Aussel}, {Lamastra}, \& {Fiore}}]{feruglio2010}
{Feruglio}, C., {Maiolino}, R., {Piconcelli}, E., {et~al.} 2010, \aap, 518,
  L155

\bibitem[{{Fischer} {et~al.}(2010){Fischer}, {Sturm}, {Gonz{\'a}lez-Alfonso},
  {Graci{\'a}-Carpio}, {Hailey-Dunsheath}, {Poglitsch}, {Contursi}, {Lutz},
  {Genzel}, {Sternberg}, {Verma}, \& {Tacconi}}]{fischer2010a}
{Fischer}, J., {Sturm}, E., {Gonz{\'a}lez-Alfonso}, E., {et~al.} 2010, \aap,
  518, L41

\bibitem[{{Fu} \& {Stockton}(2009)}]{fu2009}
{Fu}, H., \& {Stockton}, A. 2009, \apj, 690, 953

\bibitem[{{Gebhardt} {et~al.}(2000){Gebhardt}, {Bender}, {Bower}, {Dressler},
  {Faber}, {Filippenko}, {Green}, {Grillmair}, {Ho}, {Kormendy}, {Lauer},
  {Magorrian}, {Pinkney}, {Richstone}, \& {Tremaine}}]{gebhardt2000}
{Gebhardt}, K., {Bender}, R., {Bower}, G., {et~al.} 2000, \apjl, 539, L13

\bibitem[{{Greene} \& {Ho}(2006)}]{greene2006}
{Greene}, J.~E., \& {Ho}, L.~C. 2006, \apj, 641, 117

\bibitem[{{Greene} {et~al.}(2011){Greene}, {Zakamska}, {Ho}, \&
  {Barth}}]{greene2011}
{Greene}, J.~E., {Zakamska}, N.~L., {Ho}, L.~C., \& {Barth}, A.~J. 2011, \apj,
  732, 9

\bibitem[{{Hainline} {et~al.}(2011){Hainline}, {Shapley}, {Greene}, \&
  {Steidel}}]{hainline2011}
{Hainline}, K.~N., {Shapley}, A.~E., {Greene}, J.~E., \& {Steidel}, C.~C. 2011,
  \apj, 733, 31

\bibitem[{{Ho} {et~al.}(1997){Ho}, {Filippenko}, \& {Sargent}}]{ho1997}
{Ho}, L.~C., {Filippenko}, A.~V., \& {Sargent}, W.~L.~W. 1997, \apjs, 112, 315

\bibitem[{{Hopkins} {et~al.}(2006){Hopkins}, {Hernquist}, {Cox}, {Di Matteo},
  {Robertson}, \& {Springel}}]{hopkins2006}
{Hopkins}, P.~F., {Hernquist}, L., {Cox}, T.~J., {et~al.} 2006, \apjs, 163, 1

\bibitem[{{Humphrey} {et~al.}(2010){Humphrey}, {Villar-Mart{\'{\i}}n},
  {S{\'a}nchez}, {Mart{\'{\i}}nez-Sansigre}, {Delgado}, {P{\'e}rez},
  {Tadhunter}, \& {P{\'e}rez-Torres}}]{humphrey2010}
{Humphrey}, A., {Villar-Mart{\'{\i}}n}, M., {S{\'a}nchez}, S.~F., {et~al.}
  2010, \mnras, 408, L1

\bibitem[{{Husemann} {et~al.}(2013){Husemann}, {Wisotzki}, {S{\'a}nchez}, \&
  {Jahnke}}]{husemann2013}
{Husemann}, B., {Wisotzki}, L., {S{\'a}nchez}, S.~F., \& {Jahnke}, K. 2013,
  \aap, 549, A43

\bibitem[{{Kang} {et~al.}(2013){Kang}, {Woo}, {Schulze}, {Riechers}, {Kim},
  {Park}, \& {Smolcic}}]{kang2013}
{Kang}, W.-R., {Woo}, J.-H., {Schulze}, A., {et~al.} 2013, \apj, 767, 26

\bibitem[{{Komossa} \& {Schulz}(1997)}]{komossa1997}
{Komossa}, S., \& {Schulz}, H. 1997, \aap, 323, 31

\bibitem[{{Leipski} \& {Bennert}(2006)}]{leipski2006}
{Leipski}, C., \& {Bennert}, N. 2006, \aap, 448, 165

\bibitem[{{Leipski} {et~al.}(2006){Leipski}, {Falcke}, {Bennert}, \&
  {H{\"u}ttemeister}}]{leipski2006a}
{Leipski}, C., {Falcke}, H., {Bennert}, N., \& {H{\"u}ttemeister}, S. 2006,
  \aap, 455, 161

\bibitem[{{Liu} {et~al.}(2013){Liu}, {Zakamska}, {Greene}, {Nesvadba}, \&
  {Liu}}]{liu2013a}
{Liu}, G., {Zakamska}, N.~L., {Greene}, J.~E., {Nesvadba}, N.~P.~H., \& {Liu},
  X. 2013, \mnras, 430, 2327

\bibitem[{{Magorrian} {et~al.}(1998){Magorrian}, {Tremaine}, {Richstone},
  {Bender}, {Bower}, {Dressler}, {Faber}, {Gebhardt}, {Green}, {Grillmair},
  {Kormendy}, \& {Lauer}}]{magorrian1998}
{Magorrian}, J., {Tremaine}, S., {Richstone}, D., {et~al.} 1998, \aj, 115, 2285

\bibitem[{{Moe} {et~al.}(2009){Moe}, {Arav}, {Bautista}, \&
  {Korista}}]{moe2009}
{Moe}, M., {Arav}, N., {Bautista}, M.~A., \& {Korista}, K.~T. 2009, \apj, 706,
  525

\bibitem[{{Mulchaey} {et~al.}(1996){Mulchaey}, {Wilson}, \&
  {Tsvetanov}}]{mulchaey1996}
{Mulchaey}, J.~S., {Wilson}, A.~S., \& {Tsvetanov}, Z. 1996, \apjs, 102, 309

\bibitem[{{M{\"u}ller-S{\'a}nchez} {et~al.}(2011){M{\"u}ller-S{\'a}nchez},
  {Prieto}, {Hicks}, {Vives-Arias}, {Davies}, {Malkan}, {Tacconi}, \&
  {Genzel}}]{muller-sanchez2011}
{M{\"u}ller-S{\'a}nchez}, F., {Prieto}, M.~A., {Hicks}, E.~K.~S., {et~al.}
  2011, \apj, 739, 69

\bibitem[{{Nesvadba} {et~al.}(2008){Nesvadba}, {Lehnert}, {De Breuck},
  {Gilbert}, \& {van Breugel}}]{nesvadba2008}
{Nesvadba}, N.~P.~H., {Lehnert}, M.~D., {De Breuck}, C., {Gilbert}, A.~M., \&
  {van Breugel}, W. 2008, \aap, 491, 407

\bibitem[{{Netzer}(2009)}]{netzer2009}
{Netzer}, H. 2009, \apj, 695, 793

\bibitem[{{Osterbrock}(1989)}]{osterbrock1989}
{Osterbrock}, D.~E. 1989, {Astrophysics of gaseous nebulae and active galactic
  nuclei}, ed. {Osterbrock, D.~E.}

\bibitem[{{Park} {et~al.}(2012){Park}, {Woo}, {Treu}, {Barth}, {Bentz},
  {Bennert}, {Canalizo}, {Filippenko}, {Gates}, {Greene}, {Malkan}, \&
  {Walsh}}]{park2012}
{Park}, D., {Woo}, J.-H., {Treu}, T., {et~al.} 2012, \apj, 747, 30

\bibitem[{{Penston} {et~al.}(1990){Penston}, {Robinson}, {Alloin},
  {Appenzeller}, {Aretxaga}, {Axon}, {Baribaud}, {Barthel}, {Baum}, {Boisson},
  {de Bruyn}, {Clavel}, {Colina}, {Dennefeld}, {Diaz}, {Dietrich}, {Durret},
  {Dyson}, {Gondhalekar}, {van Groningen}, {Jablonka}, {Jackson},
  {Kollatschny}, {Laurikainen}, {Lawrence}, {Masegosa}, {McHardy}, {Meurs},
  {Miley}, {Moles}, {O'Brien}, {O'Dea}, {del Olmo}, {Pedlar}, {Perea}, {Perez},
  {Perez-Fournon}, {Perry}, {Pilbratt}, {Rees}, {Robson}, {Rodriguez-Pascual},
  {Rodriguez Espinosa}, {Santos-Lleo}, {Schilizzi}, {Stasi{\'n}ska}, {Stirpe},
  {Tadhunter}, {Terlevich}, {Terlevich}, {Unger}, {Vila-Vilaro}, {Vilchez},
  {Wagner}, {Ward}, \& {Yates}}]{penston1990}
{Penston}, M.~V., {Robinson}, A., {Alloin}, D., {et~al.} 1990, \aap, 236, 53

\bibitem[{{Reyes} {et~al.}(2008){Reyes}, {Zakamska}, {Strauss}, {Green},
  {Krolik}, {Shen}, {Richards}, {Anderson}, \& {Schneider}}]{reyes2008}
{Reyes}, R., {Zakamska}, N.~L., {Strauss}, M.~A., {et~al.} 2008, \aj, 136, 2373

\bibitem[{{Rupke} \& {Veilleux}(2011)}]{rupke2011}
{Rupke}, D.~S.~N., \& {Veilleux}, S. 2011, \apjl, 729, L27

\bibitem[{{Sanders} {et~al.}(1988){Sanders}, {Soifer}, {Elias}, {Madore},
  {Matthews}, {Neugebauer}, \& {Scoville}}]{sanders1988}
{Sanders}, D.~B., {Soifer}, B.~T., {Elias}, J.~H., {et~al.} 1988, \apj, 325, 74

\bibitem[{{Schlesinger} {et~al.}(2009){Schlesinger}, {Pogge}, {Martini},
  {Shields}, \& {Fields}}]{schlesinger2009}
{Schlesinger}, K., {Pogge}, R.~W., {Martini}, P., {Shields}, J.~C., \&
  {Fields}, D. 2009, \apj, 699, 857

\bibitem[{{Schmidt} \& {Green}(1983)}]{schmidt1983}
{Schmidt}, M., \& {Green}, R.~F. 1983, \apj, 269, 352

\bibitem[{{Schmitt} {et~al.}(2003){Schmitt}, {Donley}, {Antonucci},
  {Hutchings}, \& {Kinney}}]{schmitt2003}
{Schmitt}, H.~R., {Donley}, J.~L., {Antonucci}, R.~R.~J., {Hutchings}, J.~B.,
  \& {Kinney}, A.~L. 2003, \apjs, 148, 327

\bibitem[{{Sironi} \& {Socrates}(2010)}]{sironi2010}
{Sironi}, L., \& {Socrates}, A. 2010, \apj, 710, 891

\bibitem[{{Stockton} \& {MacKenty}(1987)}]{stockton1987}
{Stockton}, A., \& {MacKenty}, J.~W. 1987, \apj, 316, 584

\bibitem[{{Stoklasov{\'a}} {et~al.}(2009){Stoklasov{\'a}}, {Ferruit},
  {Emsellem}, {Jungwiert}, {P{\'e}contal}, \& {S{\'a}nchez}}]{stoklasova2009}
{Stoklasov{\'a}}, I., {Ferruit}, P., {Emsellem}, E., {et~al.} 2009, \aap, 500,
  1287

\bibitem[{{Sturm} {et~al.}(2011){Sturm}, {Gonz{\'a}lez-Alfonso}, {Veilleux},
  {Fischer}, {Graci{\'a}-Carpio}, {Hailey-Dunsheath}, {Contursi}, {Poglitsch},
  {Sternberg}, {Davies}, {Genzel}, {Lutz}, {Tacconi}, {Verma}, {Maiolino}, \&
  {de Jong}}]{sturm2011}
{Sturm}, E., {Gonz{\'a}lez-Alfonso}, E., {Veilleux}, S., {et~al.} 2011, \apjl,
  733, L16

\bibitem[{{Tremonti} {et~al.}(2007){Tremonti}, {Moustakas}, \&
  {Diamond-Stanic}}]{tremonti2007}
{Tremonti}, C.~A., {Moustakas}, J., \& {Diamond-Stanic}, A.~M. 2007, \apjl,
  663, L77

\bibitem[{{van Dokkum}(2001)}]{van-dokkum2001}
{van Dokkum}, P.~G. 2001, \pasp, 113, 1420

\bibitem[{{Vestergaard} \& {Peterson}(2006)}]{vestergaard2006}
{Vestergaard}, M., \& {Peterson}, B.~M. 2006, \apj, 641, 689

\bibitem[{{Villar-Mart{\'{\i}}n} {et~al.}(2011){Villar-Mart{\'{\i}}n},
  {Humphrey}, {Delgado}, {Colina}, \& {Arribas}}]{villar-martin2011}
{Villar-Mart{\'{\i}}n}, M., {Humphrey}, A., {Delgado}, R.~G., {Colina}, L., \&
  {Arribas}, S. 2011, \mnras, 418, 2032

\bibitem[{{Woo} {et~al.}(2006){Woo}, {Treu}, {Malkan}, \&
  {Blandford}}]{woo2006}
{Woo}, J.-H., {Treu}, T., {Malkan}, M.~A., \& {Blandford}, R.~D. 2006, \apj,
  645, 900

\bibitem[{{Woo} {et~al.}(2008){Woo}, {Treu}, {Malkan}, \&
  {Blandford}}]{woo2008}
---. 2008, \apj, 681, 925

\bibitem[{{Woo} \& {Urry}(2002)}]{woo2002}
{Woo}, J.-H., \& {Urry}, C.~M. 2002, \apj, 579, 530

\bibitem[{{Woo} {et~al.}(2005){Woo}, {Urry}, {van der Marel}, {Lira}, \&
  {Maza}}]{woo2005}
{Woo}, J.-H., {Urry}, C.~M., {van der Marel}, R.~P., {Lira}, P., \& {Maza}, J.
  2005, \apj, 631, 762

\bibitem[{{Woo} {et~al.}(2010){Woo}, {Treu}, {Barth}, {Wright}, {Walsh},
  {Bentz}, {Martini}, {Bennert}, {Canalizo}, {Filippenko}, {Gates}, {Greene},
  {Li}, {Malkan}, {Stern}, \& {Minezaki}}]{woo2010}
{Woo}, J.-H., {Treu}, T., {Barth}, A.~J., {et~al.} 2010, \apj, 716, 269

\bibitem[{{Zakamska} {et~al.}(2003){Zakamska}, {Strauss}, {Krolik}, {Collinge},
  {Hall}, {Hao}, {Heckman}, {Ivezi{\'c}}, {Richards}, {Schlegel}, {Schneider},
  {Strateva}, {Vanden Berk}, {Anderson}, \& {Brinkmann}}]{zakamska2003}
{Zakamska}, N.~L., {Strauss}, M.~A., {Krolik}, J.~H., {et~al.} 2003, \aj, 126,
  2125

\end{thebibliography}
\end{document}